\documentclass[11pt]{article}
\usepackage{amsfonts,amssymb,epsfig}
\usepackage{amsmath}
\def\sqr#1#2{{\vcenter{\hrule height.#2pt\hbox{\vrule width.#2pt
height#1pt \kern#1pt \vrule width.#2pt}\hrule height.#2pt}}}

\def\hook{\hbox{\vrule height0pt width4pt depth0.3pt
\vrule height7pt width0.3pt depth0.3pt \vrule height0pt width2pt
depth0pt} }

\newtheorem{THEOREM}{Theorem}[section]  
\newenvironment{theorem}{\begin{THEOREM} \hspace{-.85em} {\bf :} 
}%
                        {\end{THEOREM}}
\newtheorem{LEMMA}[THEOREM]{Lemma}
\newenvironment{lemma}{\begin{LEMMA} \hspace{-.85em} {\bf :} }%
                      {\end{LEMMA}}
\newtheorem{COROLLARY}[THEOREM]{Corollary}
\newenvironment{corollary}{\begin{COROLLARY} \hspace{-.85em} {\bf 
:} }%
                          {\end{COROLLARY}}
\newtheorem{PROPOSITION}[THEOREM]{Proposition}
\newenvironment{proposition}{\begin{PROPOSITION} \hspace{-.85em} 
{\bf :} }%
                            {\end{PROPOSITION}}
\newtheorem{DEFINITION}[THEOREM]{Definition}
\newenvironment{definition}{\begin{DEFINITION} \hspace{-.85em} {\bf 
:} \rm}%
                            {\end{DEFINITION}}
\newtheorem{EXAMPLE}[THEOREM]{Example}
\newenvironment{example}{\begin{EXAMPLE} \hspace{-.85em} {\bf :} 
\rm}%
                            {\end{EXAMPLE}}
\newtheorem{CONJECTURE}[THEOREM]{Conjecture}
\newenvironment{conjecture}{\begin{CONJECTURE} \hspace{-.85em} 
{\bf :} \rm}%
                            {\end{CONJECTURE}}
\newtheorem{PROBLEM}[THEOREM]{Problem}
\newenvironment{problem}{\begin{PROBLEM} \hspace{-.85em} {\bf :} 
\rm}%
                            {\end{PROBLEM}}
\newtheorem{REMARK}[THEOREM]{Remark}
\newenvironment{remark}{\begin{REMARK} \hspace{-.85em} {\bf :} 
\rm}%
                            {\end{REMARK}}
\newtheorem{CONCLUSION}[THEOREM]{Conclusion}
\newenvironment{conclusion}{\begin{CONCLUSION} \hspace{-.85em} {\bf :} 
\rm}%
                            {\end{CONCLUSION}}
\newcommand{\thm}{\begin{theorem}}
\newcommand{\lem}{\begin{lemma}}
\newcommand{\pro}{\begin{proposition}}
\newcommand{\dfn}{\begin{definition}}
\newcommand{\rem}{\begin{remark}}
\newcommand{\con}{\begin{conclusion}}
\newcommand{\xam}{\begin{example}}
\newcommand{\cnj}{\begin{conjecture}}
\newcommand{\prb}{\begin{problem}}
\newcommand{\cor}{\begin{corollary}}

\newcommand{\ethm}{\end{theorem}}
\newcommand{\elem}{\end{lemma}}
\newcommand{\epro}{\end{proposition}}
\newcommand{\edfn}{\bbox\end{definition}}
\newcommand{\erem}{\bbox\end{remark}}
\newcommand{\econ}{\bbox\end{conclusion}}
\newcommand{\exam}{\bbox\end{example}}
\newcommand{\ecnj}{\bbox\end{conjecture}}
\newcommand{\eprb}{\bbox\end{problem}}
\newcommand{\ecor}{\end{corollary}}

\newcommand{\beqn}{\begin{equation}}
\newcommand{\eeqn}{\end{equation}}

\newcommand{\bbox}{\vrule height7pt width4pt depth1pt}

\newcommand{\sect}[1]{\setcounter{equation}{0}\bigskip\medskip
\section{#1}\smallskip}
\newcommand{\subsect}[1]{\medskip\subsection{#1}\smallskip}

\def\br{\begin{eqnarray}}
\def\er{\end{eqnarray}}
\def\brn{\begin{eqnarray*}}
\def\ern{\end{eqnarray*}}
\def\er{\end{eqnarray}}
\def\eqq{&\!\!\!\!\!=\!\!\!\!\!&}
\def\eee{&\!\!\!\!\!\!\!\!\!\!&}
\def\vt{\vartheta}

\def\L{{\cal{L}}}

\def\a{\alpha}
\def\b{\beta}

\def\d{\delta}

\def\la{\lambda}

\def\r{\rho}
\def\L{\mathcal{L}}

\title{{\bf Coframe teleparallel models of gravity.\\
 Exact solutions.}\\
}
\author{
\thanks {\quad   email itin@math.huji.ac.il}
\small {\sf Yakov Itin }\\
\small {\sf Institute of Mathematics}\\
\small {\sf Hebrew University of Jerusalem}\\
\small {\sf Givat Ram, Jerusalem 91904, Israel}\\
}
\begin{document}  

\newcommand{\bi}[1]{\bibitem{#1}}
\date{}

\maketitle
\begin{abstract}
The superstring and superbrane theories include gravity as a necessary and fundamental part of a (future) unified field theory. Thus it is important to consider the alternative representations of general relativity as well as the alternative models of gravity.\\ 
We study the  coframe teleparallel theory of gravity with a most general quadratic Lagrangian. The coframe field on a differentiable manifold is a basic dynamical variable. A metric tensor as well as a metric compatible connection is generated by a coframe in a unique manner.\\
The Lagrangian is a general linear combination of   Weitzenb\"{o}ck's  quadratic invariants  with free dimensionless parameters $\r_1,\r_2,\r_3$. Every independent term of the Lagrangian is a global $SO(1,3)$-invariant 4-form. 
For a special choice of parameters which confirms with the local $SO(1,3)$ invariance  this theory gives an alternative  description of Einsteinian gravity - teleparallel equivalent of GR.\\
The field equations of the theory is studied by a ``diagonal'' coframe ansatz which is a subclass of a most general spherical-symmetric Einstein-Mayer ansatz. The restricted Lagrangian  depends only on two free parameters $\r_1,\r_3$.\\
 We obtain a formula for scalar curvature of a pseudo-Riemannian manifold with a metric constructed from the static ``diagonal'' solution of the field equation. It is proved that the sign of the scalar curvature depends only on  a relation between the parameters $\r_1$ and $\r_3$. Thus by a specific choice of free parameters a manifold of positive or negative curvature can be  obtained.
The scalar curvature vanishes only for a subclass of models with $\r_1=0$. This subclass includes  the teleparallel equivalent of GR.\\
We obtain the explicit form of all spherically symmetric static solutions of the ``diagonal'' type to the field equations for an arbitrary choice  of free parameters. We prove that the unique asymptotic-flat solution with Newtonian limit is the Schwarzschild solution that holds for a subclass of teleparallel models with $\r_1=0$. Thus the Yang-Mills-type term of the general quadratic coframe Lagrangian should be rejected.
\end{abstract}
\newpage
\sect{Introduction}  
The resent progress in superstrings/M-branes theories and recognizing the supergravity as a low-energy limit of the supersymmetry renew an interest to alternative models of classical gravity. 
For instant, the generalization of AdS/CFT correspondence due to Witten and Yau \cite {W-Y} is based on manifolds with a positive scalar curvature. The metric of such manifold can not be obtained as a  solution of the vacuum Einstein field equations which gives only the manifolds with zero scalar curvature. We will show in the sequel that the manifolds with a fixed sign of the scalar curvature are generated by generic solutions in teleparallel theories.\\
The teleparallel description of  gravity has been studied for a long time. The pioneer works of Cartan \cite {Ca},  Weitzenb\"{o}ck \cite{We} and Einstein \cite{Ei} dealt meaningful with  the various models of unified (gravity-electromagnetic) field theory. \\
The investigation in gauge field theory of gravity and in the Einstein-Cartan gravity (see  \cite{Hehl4}, \cite {Kawai1},\cite{G-H} and the references) evoked considerable interest in teleparallel geometry. In the most general metric-affine theory of gravity (MAG) \cite{{hehl95}} the teleparallel Lagrangian appears as a separated part of the general MAG Lagrangian.\\
For the recent investigations in this area see Refs. \cite{Kop} to \cite{itin4}.\\
The teleparallel approach to gravity can be viewed as a  generalization of the classical Einstein theory. The main properties of this framework are as follows:
\begin{itemize}
\item[\bf{1.}] The primary field variable of the theory (i.e. the coframe) has an intrinsic geometric sense. The basic geometrical quantities: metric, metric compatible connection and volume element can be constructed from the coframe in a unique manner.
\item[\bf{2.}] A covariant quadratic Lagrangian of the Yang-Mills type is defined in this teleparallel approach in contrast to classical general relativity (GR).
\item[\bf{3.}] A local energy-momentum current can be  defined in a covariant manner. 
\item[\bf{4.}] There exists a free-parametric wide class of possible field equations. So in order to get uniqueness additional symmetric requirements may be applied. For instance, the requirement of a symmetry under the group of local $SO(1,3)$ pseudo-rotations uniquely yields  the teleparallel equivalent of  GR.
\item[\bf{5.}] The origin of gravity in this teleparallel approach is the energy-momentum current as well as the antisymmetric spinorial current.
\item[\bf{6.}] The more general modifications of the classical theory of gravity such as metric-affine gravity (MAG) include the teleparallel Lagrangian as a self-consistent sector.
\end{itemize}
Let us give a brief account of  teleparallel geometry. Consider an $n$-dimensional differential manifold $M$ endowed with a smooth coframe field $\{\vt^a, \ a=0,..., n-1\}$. The 1-forms $\vt^a$ is declared to be pseudo-orthonormal. This assumption completely determines a metric on the manifold $M$ by the relation 
\begin{equation}\label{1-1}
g=\eta_{ab}\vt^a\otimes\vt^b,
\end{equation}
where $\eta_{ab}=(+1,-1,...,-1)$.
And conversely, the metric $g$  is a unique metric that makes the coframe $\vt^a$ pseudo-orthonormal. Thus it is possible to consider the coframe field $\vt^a$ as 
a basic dynamical variable  and to treat  the metric $g$ as only a secondary structure. 
In order to have an isotropic structure (without peculiar directions) this coframe variable should be defined only up to {\it global} pseudo-rotations. In fact the true dynamical variable is the equivalence class of coframes with the pseudo-rotation as an equivalence relation. Thus in additional to invariance under diffeomorphic transformations of the manifold $M$ the basic geometric structure has to be  global $SO(1,n-1)$ invariant. Note that the coframe field is a complex of $n^2$ independent variables in every point of $M$ while the symmetric metric tensor field has only $n(n+1)/2$ independent components. \\
An additional requirement of {\it local} $SO(1,n-1)$ invariance restricts the set of $n^2$ independent variables to a subset of $n(n+1)/2$ variables, which can be related with the independent components of the metric. In this case the coframe (teleparallel) structure coincides with the metric structure. One can still describe this metric geometry by coframes, but in this case it does not represent a different geometrical structure but provides only a different mathematical  tool - Cartan's ``Repe\'{r}e Mobile''.\\
Thus teleparallel geometry can be considered as a certain generalization of the Riemannian or pseudo-Riemannian metrical geometry. In order to have a simplest (pure)  teleparallel structure one can choose the connections on the manifold $M$ to be metric compatible. This metric compatibility of the connection is equivalent to the coframe compatibility:
\begin{equation}\label{1-2}
Dg=0 \qquad <==>\qquad D\vt^a=0,
\end{equation}
where $D$ is the covariant exterior derivative.\\ 
Thus  a 1-form of   connection ${w^a}_b$ is unique defined by the metric $g$ and also, via the relation (\ref{1-1}), by the coframe field $\vt^a$. Therefore the coframe field on the differential manifold completely defines a  certain geometrical structure - {\it coframe structure}, which can be written symbolically as
\begin{equation}
\{M,\vt^a,g(\vt^a), {w^a}_b(\vt^a)\}.\label{1-3}
\end{equation}
A straightforward generalization of this pure coframe structure can be obtained by considering the connection ${w^a}_b$ as an additional independent primary variable. Such generalization, which we write symbolically as
\begin{equation}
\{M,\vt^a,g(\vt^a), {w^a}_b\}.\label{1-4}
\end{equation}
can be refereed to as Cartan teleparallel geometry.\\
The structure can further be   generalized by taking the metric field $g$ as an independent geometrical variable: 
\begin{equation}
\{M,\vt^a,g, {w^a}_b\}.\label{1-5}
\end{equation}
This most general geometrical structure is realized in the framework of metric-affine gravity (MAG) \cite{hehl95}. \\
In the present work we study the general  teleparallel model of the type (\ref{1-3}) with a quadratic Lagrangian  on a 4D-manifold of Lorentzian  signature $\{1,-1,-1,-1\}$.  
\sect{Quadratic Lagrangian}  
The general quadratic (in the first derivatives of the coframe field $\vt^a$) Lagrangian can be written as a linear combination of the Weitzenb\"{o}ck  second order  invariants \cite{We}. This Lagrangian is known as a translation invariant Lagrangian  of Rumpf \cite{rumpf} :
\begin{equation}
\L=\frac{1}{2\ell^2}\sum^3_{i=1}a_i(d\vt^a\wedge*{}^{(i)}d\vt_a),
\label{2-1}
\end{equation}
where $\ell$ is the Plank length and $*$ is the Hodge dual operator. 
\footnote{Here and later the down-indexed coframe is obtained by the Lorentzian metric, thus $\vt^0=\vt_0, \ \vt^i=-\vt_i$, where $ i=1,2,3$.}\\
The Lagrangian includes the strength $d\vt^a$ and its irreducible (under the global Lorentz group) parts ${}^{(i)}d\vt_a$. The explicit expressions for these irreducible parts are: 
\br
\label{2-2}
{}^{(1)}d\vt^a&=&d\vt^a-{}^{(2)}d\vt^a-{}^{(3)}d\vt^a,\\
\label{2-3}
{}^{(2)}d\vt^a&=&\frac 13 \vt^a\wedge(e_b\hook d\vt^b),\\
\label{2-4}
{}^{(3)}d\vt^a&=&\frac 13 e^a\hook (\vt^b\wedge d\vt_b).
\er
The coefficients $a_1,a_2,a_3$ in (\ref{2-1}) are dimensionless constants. 
 Rearranging the different parts of the Lagrangian (\ref{2-1}) results in a linear combination of quadratic terms  
\begin{equation}\label{2-5}
\L=\frac{1}{2\ell^2}\sum_{i=1}^3 \rho_{i} \; {}^{(i)}V\
\end{equation}
with 
\br\label{2-6}
{}^{(1)}V &=&d\vt^a \wedge *d\vt_a,\\
\label{2-7}
{}^{(2)}V &=&\Big(d\vt_a \wedge \vt^a \Big) \wedge*\Big(d\vt_b\wedge\vt^b\Big), \\
\label{2-8}
{}^{(3)}V &=&\left(d\vt_a \wedge\vt^b \right) \wedge *\Big(d\vt_b \wedge \vt^a \Big).
\er 
The dimensionless coefficients $a_i$ and $\r_i$ are related \cite{Hehl98} as: 
\begin{align}
\label{2-9} \rho_{1} & = \frac{1}{3} \, \left(a_2 + 2 a_1 \right)  \;,& 
\rho_{2} & = \frac{1}{3} \, \left(a_3 - a_1 \right) \;,&
\rho_{3} & = \frac{1}{3} \, \left(a_1 - a_2 \right) \;,\\ 
\label{2-10} a_1 & = \rho_{1} + \rho_{3} \;,&
a_2 & = \rho_{1} - 2 \rho_{3} \;, &
a_3 & = \rho_{1} + 3 \rho_{2} + \rho_{3} \;.
\end{align}
Every independent part of the Lagrangians (\ref{2-1}) and (\ref{2-5}) is diffeomorphic covariant and invariant under the global (rigid) Lorentz group - translation invariant. Thus different choices of the free parameters $\a_i$ or $\r_i$ yield different translation invariant models of gravity.\\
Let us list the $a_i$ and the $\rho_i$ coefficients for  different teleparallel Lagrangians \cite{Hehl98}:
\begin{center}
\begin{tabular}{c|rrrrrr}
  & $\text{GR}_\parallel$ \cite{G-H}& vdH \cite{heyde76} & viable
  & YM &YM$^\dagger$& $\text{KI}$ \cite{K-I}\\ \hline $a_1$ &
  1 & 1 & 1 & 1 & 0 & 1\\ $a_2$ & $-2$ & $-2$ &$-2$  & 1 &3 & 4 \\ 
  $a_3$ & $-\frac{1}{2}$ & 1 &\text{arb.}  & 1& 0& 1\\ \hline $\rho_1$
  & 0 & 0 & 0 & 1 &1& 2\\ $\rho_2$ & $-\frac{1}{2}$ & 0 & \text{arb.} & 0 &0&
  0\\ $\rho_3$ & 1 & 1 & 1 & 0&$-1$&$-1$
\end{tabular}
\end{center}
Note also the following possibility to modify the Lagrangians (\ref{2-1}) and (\ref{2-5}). 
One can add to them a zeroth part 
\begin{equation}
\label{2-11}
\eta=\frac 14\Lambda\vt^a\wedge*\vt_a.
\end{equation}
It is easy to see that this term is an analog of the cosmological $\Lambda$-term in Einsteinian gravity. We will not use  this term in the sequel.
\sect{Field equation and ansatz}  
In order to obtain the field equation for the coframe field $\vt^a$ one has to calculate the variation of the Lagrangians (\ref{2-1}) or (\ref{2-5})  relative to a free variation of the coframe. This procedure is not quite regular because of the Hodge operator $*$, which itself depends on the coframe field $\vt^a$ and should also be a subject of variation. Thus we have, in general, for an arbitrary form $w$, a non-commutative relation
 \begin{equation}\label{3-1}
\d*w-*\d w\ne 0.
\end{equation}
The variation of the Lagrangians (\ref{2-1}), (\ref{2-5}) can be produced in the following ways:\\
1) Directly by using the Euler-Lagrange equations \cite{gron97}
 \begin{equation}
\label{3-2}
d\Big(\frac{\partial V}{\partial d\vt^a}\Big)+\frac{\partial V}{\partial \vt^a}=0.
\end{equation}
2) By using the master formula of \cite {Hehl98} - the explicit expression for the commutator (\ref{3-1}).\\
3) By using  a formula \cite{it4} for variation  of the quadratic Lagrangians of the form
$\d(\a\wedge *\b)$.\\
All these different methods of variation result in the same  second order field equation, that we will write symbolically as 
\br\label{3-3}
&&\rho_1\Big(2{}^{(1)}L_a+{}^{(1)}Q_a-2{}^{(2)}Q_a\Big)+\rho_2\Big(-2{}^{(2)}L_a+2{}^{(3)}Q_a+{}^{(4)}Q_a-2{}^{(5)}Q_a\Big)+\nonumber\\
&&\rho_3\Big(-2{}^{(3)}L_a+2{}^{(6)}Q_a+{}^{(7)}Q_a-2{}^{(8)}Q_a\Big)=0,
\er
where the leading part (which includes all the pieces with  the second order derivatives) is a linear combination  of the terms:
\br
\label{3-4}
{}^{(1)}L_a&=&d*d\vt_a,\\
\label{3-5}
{}^{(2)}L_a&=&\vt_a \wedge d *(d\vt^b\wedge \vt_b),\\
\label{3-6}
{}^{(3)}L_a&=&\vt_b \wedge d*( \vt_a \wedge d \vt^b ).
\er
The quadratic part (which involves only the pieces quadratic in the first derivatives) is a combination of the terms
\br
\label{3-7}
{}^{(1)}Q_a&=&e_a\hook(d\vt^b\wedge*d\vt_b),\\ 
\label{3-8}
{}^{(2)}Q_a&=&(e_a\hook d\vt^b)\wedge*d\vt_b,\\ 
\label{3-9}
{}^{(3)}Q_a&=&d\vt_a \wedge * ( d\vt^b \wedge \vt_b),\\ 
\label{3-10}
{}^{(4)}Q_a&=&e_a\hook\Big(d\vt^c\wedge\vt_c\wedge*(d\vt^b\wedge\vt_b)\Big),\\ 
\label{3-11}
{}^{(5)}Q_a&=&(e_a\hook d\vt^b)\wedge\vt_b\wedge*(d\vt^c\wedge\vt_c),\\ 
\label{3-12}
{}^{(6)}Q_a&=&d\vt_b\wedge*(\vt_a\wedge d\vt^b),\\ 
\label{3-13}
{}^{(7)}Q_a&=&e_a\hook\Big(\vt_c\wedge d\vt^b\wedge*(d\vt^c\wedge\vt_b)\Big),\\ 
\label{3-14}
{}^{(8)}Q_a&=&(e_a\hook d\vt^b)\wedge\vt_c\wedge*(d\vt^c\wedge\vt_b ).
\er
We are mostly interested in  static, spherical symmetric solutions of the equation (\ref{3-1}). It is well known that a general static spherical symmetric metric can be written as 
\begin{equation}\label{3-15}
ds^2=e^{2f}dt^2-e^{2g}(dx^2+dy^2+dz^2),
\end{equation}
where $f=f(r)$, $g=g(r)$.\\
Due to Einstein-Mayer  \cite{E-M} a general static, spherical symmetric coframe has the form
\begin{equation}\label{3-16}
\vt^0=e^fdt, \qquad \vt^i=wx^idt+e^gdx^i, \qquad i=1,2,3,
\end{equation}
where $f=f(r)$, $g=g(r)$ and $w=w(r)$.\\
We begin the consideration  with a simplified ``diagonal'' ansatz
\begin{equation}\label{3-17}
\vt^0=e^{f(x,y,z)}dt, \qquad \vt^i=e^{g(x,y,z)}dx^i, \qquad i=1,2,3.
\end{equation}
and in a sequel we will study the spherical-symmetric ``diagonal'' solutions to the field equation (\ref{3-3}).
Thus we restrict our attention to a sub-family of solutions with an identically vanishing function $w$. This restriction is mainly made in order to simplify the calculations. The consideration of a non-restricted spherical-symmetric coframe as well as the study of the physical sense of its ``non-diagonal'' part will be made in a sequel paper. Note, however, that even the restricted coframe leads to a most general spherical-symmetric metric (\ref{3-15}). Thus it can be believed that the  ``non-diagonal'' pieces of the coframe can be related to an additional (possible, non-gravity) interaction. \\ 
For the ``diagonal'' coframe (\ref{3-17}) the following ``parallel'' relation holds
\begin{equation}\label{3-18}
\vt_a\wedge d\vt^a=0. 
\end{equation}
This condition is diffeomorphic and global $SO(1,3)$ invariant, thus it can serve as a necessary condition to have a ``diagonal'' coframe. 
The relation (\ref{3-18}) shows that the multiplier of the coefficient $\rho_2$ in the Lagrangian (\ref{2-5}) is identically zero. Therefore the field equation (\ref{3-3}) reduces to:
\br\label{3-19}
&&\rho_1\Big(2{}^{(1)}L_a+{}^{(1)}Q_a-2{}^{(2)}Q_a\Big)+\rho_3\Big(-2{}^{(3)}L_a+2{}^{(6)}Q_a+{}^{(7)}Q_a-2{}^{(8)}Q_a\Big)=0.
\er
The coframe (\ref{3-17}) being substituted (Appendix A) in the field equation (\ref{3-19}) splits it in the temporal ($a=0$) and in the spatial ($a=i=1,2,3$) parts.  
The temporal part  takes the form 
\begin{equation}\label{3-20}
\rho_1\Big(-2\triangle f-2(\nabla f\cdot\nabla g)+2(\nabla g)^2-(\nabla f)^2\Big)*\vt^0 +\rho_3\Big(-4\triangle g-2(\nabla g)^2\Big)*\vt^0  =0,
\end{equation}
which is equivalent to a scalar equation
\begin{equation}\label{3-21}
\rho_1\Big(2\triangle f+2(\nabla f\cdot\nabla g)-2(\nabla g)^2+(\nabla f)^2\Big)+2\rho_3\Big(2\triangle g+(\nabla g)^2\Big)=0.
\end{equation}
As for the spatial part of the field equation  (\ref{3-19}) 
\br\label{3-22}
&&\!\!\!\!\!\!\rho_1\bigg(\Big(2\triangle g+2(\nabla f\cdot\nabla g)-(\nabla f)^2\Big)\eta_{ik}+2g_{ik}+2f_ig_k-2f_if_k-2g_ig_k\bigg)*\vt^k+\nonumber\\
&&\!\!\!\!\!\!2\rho_3\bigg(\Big(\triangle f+\triangle g+(\nabla f)^2\Big)\eta_{ik}+ f_{ik}+g_{ik}+f_if_k-f_ig_k-g_if_k-g_ig_k\bigg)*\vt^k=0.
\er
The system (\ref{3-20}) and (\ref{3-22}) is exhibited in the scalar form by 
\begin{equation}\label{3-23}
\left\{
\begin{array}{l}
\rho_1\Big(2\triangle f+2(\nabla f\cdot\nabla g)-2(\nabla g)^2+(\nabla f)^2\Big)+2\rho_3\Big(2\triangle g+(\nabla g)^2\Big)=0\\
\rho_1\bigg(\Big(2\triangle g+2(\nabla f\cdot\nabla g)-(\nabla f)^2\Big)\eta_{ik}+2g_{ik}+2f_ig_k-2f_if_k-2g_ig_k\bigg)\\
\qquad +2\rho_3\bigg(\Big(\triangle f+\triangle g+(\nabla f)^2\Big)\eta_{ik}+ f_{ik}+g_{ik}+f_if_k-f_ig_k-g_if_k-g_ig_k\bigg)=0.
\end{array}\right.
\end{equation}
The last equation is symmetric in respect to transposition  of the indices 
($i<\!\!-\!\!>j$). Thus the result is  an over-determined  system  of 11 equations for two functions $f$ and $g$. This system certainly has a trivial solution $f=g=0$ which yields a flat Minkowskian metric. The interesting fact is that this over-determined system has  also nontrivial solutions for an almost all values of the free parameters $\r_1,\r_3$. The explicit form of these solutions will be obtained in the sequel.  

\sect{Curvature}              
In this section we establish a formula for  scalar curvature of a manifold, which metric is constructed from the solution of the system (\ref{3-23}). The result does not depend on a specific solution of the field equation.    For actual calculation of the scalar curvature we use the formulas  \cite {L_L} for non-vanishing components of Ricci tensor in the case 
of ``diagonal'' metrics.\\
Let the components of the metric tensor be  
\footnote{In all formulas in this section the summation over repeated 
indices is not used.}
\begin{equation}\label{4-1}
g_{\a\a}=e_\a e^{2F_\a}, \qquad g_{\a\b}=0 \qquad  {\text{for}} \qquad \a\ne \b,
\end{equation}
where $e_\a=\pm 1$.\\
The components of Ricci tensor are:
\begin{equation}\label{4-2}
R_{\a\b}=\sum_{\mu\ne \a,\b}\Big(F_{\mu,\b}F_{\b,\a}+F_{\a,\b}F_{\mu,\a}-F_{\mu,\a}F_{\mu,\b}-F_{\mu,\a,\b}\Big) \qquad {\text{for}} \qquad \a\ne \b,
\end{equation}
\br\label{4-3}
R_{\a\a}&=&\sum_{\mu\ne \a}\Big [F_{\a,\a}F_{\mu,\a}-F_{\mu,\a}^2-F_{\mu,\a,\a} \nonumber\\
&+&e_\a e_\mu e^{2(F_\a-F_\mu)}\Big(F_{\mu,\mu}F_{\a,\mu}-F_{\a,\mu}^2-F_{\a,\mu,\mu}-F_{\a,\mu}\sum_{\nu\ne \a,\mu}F_{\nu,\mu}\Big)\Big].
\er
For  the diagonal coframe (\ref{3-17})  we have
\begin{equation}\label{4-4}
e_0=-1,\quad e_i=1,\qquad F_0=f,\quad F_i=g \quad\text{for}\quad i=1,2,3.
\end{equation}
Hence using (\ref{4-3}) we obtain
\brn
R_{00}&=&-e^{2(f-g)}\bigg(\Big(F_{1,1}F_{0,1}-F_{0,1}^2-F_{0,1,1}-F_{0,1}(F_{2,1}+F_{3,1})\Big)+\\
&&\qquad \qquad\Big(F_{2,2}F_{0,2}-F_{0,2}^2-F_{0,2,2}-F_{0,2}(F_{1,2}+F_{3,2})\Big)+\\
&&\qquad \qquad \Big(F_{3,3}F_{0,3}-F_{0,3}^2-F_{0,3,3}-F_{0,3}(F_{1,3}+F_{2,3})\Big)\bigg)\\
&=&e^{2(f-g)}\Big(f_1^2+f_{11}+f_1g_1+f_2^2+f_{22}+f_2g_2+f_3^2+f_{33}+f_3g_3\Big).
\ern
Thus the temporal component of the Ricci tensor is
\begin{equation}\label{4-5}
R_{00}=e^{2(f-g)}\Big(\triangle f+(\nabla f)^2+(\nabla f\cdot\nabla g)\Big).
\end{equation}
Using (\ref{4-2}) we compute
\brn
R_{11}\eqq\Big [F_{1,1}F_{0,1}-F_{0,1}^2-F_{0,1,1} 
+e_1 e_0 e^{2(F_1-F_0)}\Big(F_{0,0}F_{1,0}-F_{1,0}^2-F_{1,0,0}-F_{1,0}(F_{2,0}+F_{3,0})\Big)\Big]+\\
\eee \Big [F_{1,1}F_{2,1}-F_{2,1}^2-F_{2,1,1}
+e_1 e_2 e^{2(F_1-F_2)}\Big(F_{2,2}F_{1,2}-F_{1,2}^2-F_{1,2,2}-F_{1,2}
(F_{0,2}+F_{3,2})\Big)\Big]+\\
\eee \Big [F_{1,1}F_{3,1}-F_{3,1}^2-F_{3,1,1} 
+e_1 e_3 e^{2(F_1-F_3)}\Big(F_{3,3}F_{1,3}-F_{1,3}^2-F_{1,3,3}-F_{1,3}(F_{0,3}+F_{2,3})\Big)\Big]\\
\eqq\Big [g_1f_1-f_1^2-f_{11}\Big]-
\Big [g_{11}+\Big(g_{22}+g_2(f_2+g_2)\Big)\Big]-
\Big [g_{11}+\Big(g_{33}+g_3(f_3+g_3)\Big)\Big].
\ern
Thus
\begin{equation}\label{4-6}
R_{11}=-f_{11}-g_{11}+2g_1f_1-f_1^2+g_1^2-\triangle g-(\nabla g\cdot \nabla f)-(\nabla g)^2.
\end{equation}
Therefore, the spatial components of the Ricci curvature tensor are
\begin{equation}\label{4-7}
R_{ii}=-f_{ii}-g_{ii}+2g_if_i-f_i^2+g_i^2-\triangle g-(\nabla g\cdot \nabla f)-(\nabla g)^2.
\end{equation}
Consequently, the scalar curvature takes the form
\begin{equation}\label{4-8}
\boxed{R=-e^{-2g}\Big(\triangle f+2\triangle g+(\nabla f)^2+(\nabla f\cdot\nabla g)+(\nabla g)^2\Big).}
\end{equation}
Note that this formula is valid for the diagonal coframe (\ref{3-17}) even if it does not satisfy the field equation (\ref{3-23}). Calculate now the scalar curvature of the pseudo-Riemannian manifold constructed from a solution of the field equation. Taking the trace of the second equation of the system (\ref{3-23}) and adding the first equation we obtain
 \begin{equation}\label{4-9}
2\r_1\Big(2\triangle g+\triangle f+3(\nabla f\cdot\nabla g)\Big)+4\r_3\Big(2\triangle g+\triangle f+(\nabla f)^2+(\nabla f\cdot\nabla g)+(\nabla g)^2\Big)=0.
\end{equation}
Substituting the second derivatives expression from the formula (\ref{4-8}) we get
\begin{equation}\label{4-10}
Re^{2g}(\r_1+2\r_3)=-\r_1(\nabla f-\nabla g)^2.
\end{equation}
We will see later that in the special case $\r_1+2\r_3=0$ the system (\ref{3-23}) has a non-trivial solution only if  $f=g$. \\
Thus the sign of the scalar curvature depends only on the relation between free parameters $\r_1,\r_3$ and does not changes on the manifold. This sign is an invariant  for every particular teleparallel model (particular choice of the parameters $\r_1,\r_3$). \\ 
The conclusion is:
\pro
Suppose that $\r_1+2\r_3\ne0$. The ``diagonal'' solutions of the field equation (\ref{3-3})  describe the pseudo-Riemannian manifold with the scalar curvature 
\begin{equation}\label{4-11}
\boxed{R=-\frac {\r_1}{\r_1+2\r_3}e^{-2g}(\nabla f-\nabla g)^2}
\end{equation}
 The sign of the scalar curvature $R$ is opposite to the sign of $\r_1(\r_1+2\r_3)$ and it is an invariant of a particular teleparallel model.\\
The scalar curvature  is zero if and only if $\r_1=0$. - the teleparallel equivalent of GR.
\epro
The case $\r_1+2\r_3=0$ will be treated later. Note that the subclass $\r_1=0$ includes the Einstein theory of gravity in the form of teleparallel equivalent of GR.\\
We have shown that the scalar curvature of manifold constructed from a ``diagonal'' solution of a specific teleparallel model has a fixed sign.  In order to clarify the result let us recall a similar fact from the geometry of surfaces. Consider, as an example, a 2D-surface with a metric 
\begin{equation}\label{4-12}
dl^2=e^{2f}(du^2+dv^2),
\end{equation} 
where $f=f(u,v)$.
Note that for an arbitrary 2D-manifold every metric can be transformed to this conformal invariant form. 
The Gauss curvature for the metric (\ref{4-12}) takes the form
\begin{equation}\label{4-13}
K=- e^{-2f}\triangle f.
\end{equation} 
Consider now a class of metrics that satisfy a ``covariant field equation'':
\begin{equation}\label{4-14}
\triangle f=\rho (\nabla f)^2.
\end{equation} 
The  Gauss curvature takes now the form
\begin{equation}\label{4-15}
K=-{\rho}\Big(e^{-2f}(\nabla f)^2\Big).
\end{equation} 
The term in the brackets is positive, thus the sign of the curvature depends only of a free parameter $\rho$ of the model. Note that (\ref{4-15}) is closely similar to the formula (\ref{4-11}).
The curvature vanishes only in the case $\rho=0$, i.e. $f$ is a harmonic function.  This is a 2D Euclidean analog of Einstein gravity.\\
In fact we have proved this property of having fixed sign only in a case of ``diagonal'' metric. \\
Our conjecture that this property can be proved  for an arbitrary static metric that satisfies the general teleparallel field equation.   
\sect{Spherical-symmetric ansatz}   
Consider a solution of the field equation which depends only on one radial variable $s=x^2+y^2+z^2$.
\begin{equation}\label{5-1}
\vt^0=e^{f(s)}dt^0,\qquad \vt^i=e^{g(s)}dt^i,\qquad i=1,2,3.
\end{equation}
With this ansatz we reduce the general problem of determining 16 unknown functions (the components of the coframe $\vt^a$) of four variables (coordinates) to determining two functions $f$ and $g$ of one radial variable $s$. 
We obtain, instead of (\ref{3-23}), an over-determined system of three second order ODE for two independent variables 
\begin{equation}\label{5-2}
\left\{
\begin{array}{l}
\rho_1\Big(2f''s+3f'+2f'g's-2(g')^2s+(f')^2s\Big)+
2\rho_3\Big(2g''s+3g'+(g')^2s\Big)=0\\
\rho_1\Big(2g''+2f'g'-2(f')^2-2(g')^2\Big)+
2\rho_3\Big(f''+g''+(f')^2-2f'g'-(g')^2\Big)=0\\
\rho_1\Big(4g''s+4g'+4f'g's-2(f')^2s\Big)+
2\rho_3\Big(2f''s+2f'+2g''s+2g'+2(f')^2s\Big)=0.
\end{array}\right.
\end{equation}
We will see later that for certain special values of the free parameters $\r_1,\r_3$ these three equations are algebraically dependent and the system (\ref{5-2}) reduces to a good posed system of two independent equations for two independent variables. For most of the values of free parameters  $\r_1,\r_3$ the three equations of (\ref{5-2}) are independent. Using this fact we obtain an explicit form of exact solutions. Let us make an algebraic transformation of the system (\ref{5-2}), which allow to decrease the order of the system. 
The subtraction of the second equation of (\ref{5-2}) multiplied by $2s$ from the third equation multiplied by $3$ results in
\brn
&&\rho_1\Big(4g''s+6g'+4f'g's-(f')^2s+2(g')^2s\Big)+\\
&& \qquad \qquad 2\rho_3\Big(2f''s+3f'+2g''s+3g'+2(f')^2s+2f'g's+ (g')^2s\Big)=0.
\ern
The subtraction of the second equation of (\ref{5-2}) multiplied by $(2s)$  from the third equation  results in
$$
\rho_1\Big(2g'+(f')^2s+2(g')^2s\Big)+
2\rho_3\Big(f'+g'+2f'g's+(g')^2s\Big)=0.
$$
Thus we obtain an equivalent system to (\ref{5-2})
\begin{equation}\label{5-3}
\left\{
\begin{array}{l}
\rho_1\Big(2f''s+3f'+2f'g's-2(g')^2s+(f')^2s\Big)+
2\rho_3\Big(2g''s+3g'+(g')^2s\Big)=0\\
\rho_1\Big(4g''s+6g'+4f'g's-(f')^2s+2(g')^2s\Big)+\\
\qquad \qquad \qquad \qquad 2\rho_3\Big(2f''s+3f'+2g''s+3g'+2(f')^2s+2f'g's+(g')^2s\Big)=0\\
\rho_1\Big(2g'+(f')^2s+2(g')^2s\Big)+
2\rho_3\Big(f'+g'+2f'g's+(g')^2s\Big)=0
\end{array}\right.
\end{equation}
We can now decrease the order of the system (\ref{5-3}) by introducing new variables
\begin{equation}\label{5-4}
u=f's^{\frac 32},\qquad v=g's^{\frac 32}.
\end{equation}
WE obtain  a system of first order ODE
\begin{equation}\label{5-5}
\left\{
\begin{array}{l}
\rho_1\Big(2s^{\frac 32}u'+2uv-2v^2+u^2\Big)+
2\rho_3\Big(2s^{\frac 32}v'+v^2\Big)=0\\
\rho_1\Big(4s^{\frac 32}v'+4uv-u^2+2v^2\Big)+
2\rho_3\Big(2s^{\frac 32}u'+2s^{\frac 32}v'+2u^2+2uv+v^2\Big)=0\\
\rho_1\Big(2vs^{\frac 12}+u^2+2v^2\Big)+
2\rho_3\Big(us^{\frac 12}+vs^{\frac 12}+2uv+v^2\Big)=0.
\end{array}\right.
\end{equation}
Note that the first two equations of (\ref{5-5}) contain the first order derivatives and the quadratic expressions of the ``strength'' variables $u$ and $v$. This is a general form of a non-linear (Einstein type) equation in gravity. The third equation of (\ref{5-5}) is an algebraic relation between the variables $u$ and $v$ and the radial coordinate $s$. \\ 
Solving the first two equations of (\ref{5-5}) for the derivatives terms results in
\begin{equation}\label{5-6}
\left\{
\begin{array}{l}
\!\!(2\r_3+\r_1)(\r_3-\r_1)\Big(2s^{\frac 32}u'\Big)\!\!=\!\!(\r_1^2+2\r_3\r_1-4\r_3^2)u^2-2(2\r_3^2+\r_1\r_3-\r_1^2)uv-2\r_1(\r_3+\r_1)v^2\\
\!\!2(2\r_3+\r_1)(\r_3-\r_1)\Big(2s^{\frac 32}v'\Big)\!=\!\r_1(2\r_3-\r_1)u^2+4\r_1^2uv-2(2\r_3^2-3\r_1\r_3-\r_1^2)v^2
\\ 
\!\!\rho_1\Big(2vs^{\frac 12}+u^2+2v^2\Big)+2\rho_3\Big(us^{\frac 12}+vs^{\frac 12}+2uv+v^2\Big)=0.
\end{array}\right.
\end{equation}
The three equations (\ref{5-5}) or (\ref{5-6}) contain the derivatives $u', v'$ so one can hope to obtain an algebraic equations for the functions $u$ and $v$. In order to obtain such algebraic relation we writhe the second and the third equations of the system (\ref{5-5}) as 
\br\label{5-7}
&&(\r_1v'+\r_3u'+r_3v')=Fs^{-3/2},\\
\label{5-8}
&&(\r_1v+\r_3u+r_3v)=Gs^{-1/2},
\er
where the quadratic functions of $u$ and $v$ are
\br\label{5-9}
F&=&F(u,v)=-\frac 14 \Big((4\r_3-\r_1)u^2+4(\r_1+\r_3)uv+2(\r_1+\r_3)v^2\Big),\\
\label{5-10}
G&=&G(u,v)=-\frac 12 \Big(\r_1u^2+4\r_3uv+2(\r-1+\r_3)v^2\Big).
\er
Taking the derivative of the (\ref{5-8}) and comparing it with (\ref{5-7}) we obtain
\begin{equation}\label{5-11}
\frac{\partial G}{\partial u}u'+\frac{\partial G}{\partial v}v'=\Big(F+\frac 12 G\Big)s^{-1}.
\end{equation}
The derivative terms in (\ref{5-6}) can be written symbolically as 
\begin{equation}\label{5-12}
u'=\phi_1(u,v)s^{-3/2}, \qquad v'=\phi_2(u,v)s^{-3/2}.
\end{equation}
Substituting this equations in (\ref{5-11}) we obtain an algebraic equation that steel contains the independent variable $s$
\begin{equation}\label{5-13}
\frac{\partial G}{\partial u}\phi_1+\frac{\partial G}{\partial v}\phi_2=\Big(F+\frac 12 G\Big)s^{1/2}.
\end{equation}
Deriving now $s$ from (\ref{5-8}) and substituting in (\ref{5-13}) we obtain an algebraic equation for $u$ and $v$ which we are looking for 
\begin{equation}\label{5-14}
(\r_1v+\r_3u+\r_3v)\bigg(\frac{\partial G}{\partial u}\phi_1+\frac{\partial G}{\partial v}\phi_2\bigg)=\Big(F+\frac 12 G\Big)G.
\end{equation}
In order to write this equation in the explicit form we make a use of an algebraic computations program ``REDUCE''. The result is
\begin{equation}\label{5-15}
\r_1(\r_1-\r_3)(u-v)(\r_1v+\r_3u+\r_3v)\Big(\r_1u^2+4\r_3uv+2(\r_1+\r_3)v^2\Big)=0.
\end{equation}
This equation  gives us a necessary condition to obtain a solution of the system (\ref{5-5}) or (\ref{5-6}).
\pro
In order to have a  nontrivial spherical-symmetric static ``diagonal'' solution 
of the field equation (\ref{3-3})  one of the following condition should be  satisfied:
\begin{itemize}
\item[1.] The free parameters satisfy the relation 
\begin{equation}\label{5-16}
\r_1(\r_1-\r_3)=0.
\end{equation}
\item[2.] Function $u$ is zero thus 
\begin{equation}\label{5-17}
\r_1+\r_3=0.
\end{equation}
\item[3.] Function $v$ is zero thus 
\begin{equation}\label{5-18}
\r_3=0.
\end{equation}
\item[4.] The nonzero functions $u$ and $v$ satisfy the proportional relation
\begin{equation}\label{5-19}
u=\la v, \qquad \la=\la(\r_1,\r_3).
\end{equation}
\end{itemize}
\epro
\sect{Solutions}              
Now we are able to derive the explicit form of the solutions for the system (\ref{5-2}) for different values of the free parameters $\r_1,\r_3$. 
It is useful to write the formula (\ref{4-11}) for the scalar curvature via the radial functions $u$ and $v$.
Using the relations
$$\nabla f^2=4(f')^2s,\qquad \nabla g^2=4(g')^2s,\qquad \text{and}\qquad\nabla f\cdot\nabla g=4f'g's.$$
we obtain
\brn
R&=&-\frac {\r_1}{\r_1+2\r_3}e^{-2g}(\nabla f-\nabla g)^2=
-4\frac {\r_1}{\r_1+2\r_3}e^{-2g}s(f'-g')^2.
\ern
Thus the scalar curvature is 
\begin{equation}\label{6-1}
R=-\frac 4{s^2}\frac {\r_1}{\r_1+2\r_3}e^{-2g}(u-v)^2.
\end{equation}
\subsect{GR-type Lagrangian: \ $\boldsymbol{\r_1=0}$}
The system (\ref{5-5}) takes the form
\begin{equation}\label{6-2}
\left\{
\begin{array}{l}
2s^{\frac 32}v'+v^2=0\\
2s^{\frac 32}u'+2s^{\frac 32}v'+2u^2+2uv+v^2=0\\
us^{\frac 12}+vs^{\frac 12}+2uv+v^2=0
\end{array}\right.
\end{equation}
Integrate the first equation to obtain
\begin{equation}\label{6-3}
v=-\frac{\sqrt{s}}{c\sqrt{s}+1},
\end{equation}
where $c$ is a constant of integration.\\
Substituting this relation in the third equation of the system (\ref{6-2}), we find
\begin{equation}\label{6-4}
u=\frac{cs}{c^2s-1}.
\end{equation}
Substituting the functions (\ref{6-3} - \ref{6-4})  in the second equation of (\ref{6-2}) we obtain an identity.\\
Integrate
\footnote{The constant of integration on this step can be omitted, because we can still re-scale the coordinates.}
the functions $u$ and $v$ to obtain the ansatz functions ($r=\sqrt s$):
\begin{equation}\label{6-5}
f=\ln\frac{1-\frac 1{cr}}{1+\frac 1{cr}},\qquad g=2\ln{(1+\frac 1{cr})}.
\end{equation}
By taking the parameter of integration to be inversely proportional to the mass of the central body $c=\frac 2m$  we obtain the coframe field in the form
\br\label{6-6}
\vt^0&=&\frac{1-\frac m{2r}}{1+\frac m{2r}}dt,\qquad \vt^i=(1+\frac m{2r})^2dx^i, \qquad i=1,2,3.
\er
This coframe field yields the Schwarzschild metric in isotropic coordinates
\begin{equation}\label{6-7}
ds^2=\bigg(\frac{1-\frac m{2r}}{1+\frac m{2r}}\bigg)^2dt^2-\Big(1+\frac m{2r}\Big)^4(dx^2+dy^2+dz^2).
\end{equation}
Note that the parameter $\r_2$ is not determined via the ``diagonal'' ansatz. Thus the Schwarzschild metric is a solution for a family of teleparallel field equations which defined by the parameters: $\r_1=0$ and $ \r_2,\r_3$ - arbitrary. Note that for the teleparallel equivalent of GR we need beside of zero $\r_1$  an additional condition: $\r_3+2\r_2=0$ .
\subsect{$\boldsymbol{\r_1=\r_3}$}
The system (\ref{5-5}) takes the form
\begin{equation}\label{6-8}
\left\{
\begin{array}{l}
2s^{\frac 32}(u'+2v')+u(u+2v)=0\\
4s^{\frac 32}(u'+2v')+(u+2v)(3u+2v)=0\\
2s^{\frac 12}(u+2v)+(u+2v)^2=0.
\end{array}\right.
\end{equation}
Subtracting the first equation (multiplied by 2) from the second one we obtain that the system is satisfied if and only if 
\begin{equation}\label{6-9}
u+2v=0, \qquad ==>\qquad f+2g=0.
\end{equation}
Consequently the coframe is 
\begin{equation}\label{6-10}
\vt^0=e^{-2g(r)}dt, \qquad \vt^i=e^{g(r)}dx^i
\end{equation}
and the metric is 
\begin{equation}\label{6-11}
ds^2=e^{-4g(r)}dt^2-e^{2g(r)}(dx^2+dy^2+dz^2).
\end{equation}
Thus the solution incorporates an unknown function $g=g(r)$.
The scalar curvature of the metric (\ref{6-11}) is negative for an arbitrary choice of the function $g=g(r)$.
\begin{equation}\label{6-12}
R=-3(g')^2e^{-2g}.
\end{equation}
\subsect{$\boldsymbol{ \ u=0\qquad ==>\qquad \r_1+\r_3=0}$}
The equation (\ref{5-15}) can have a non-trivial solution  with a zero function $u$ if and only if the parameters satisfy 
\begin{equation}\label{6-13}
\r_1+\r_3=0.
\end{equation}
This is a case of $YM^\dagger$ Lagrangian 
\begin{equation}\label{6-14}
\L=d^\dagger\vt^a\wedge *d^\dagger\vt_a,
\end{equation}
where $d^\dagger$ is a coderivative operator.\\
The second and the third equations of (\ref{5-5}) are satisfied identically for zero $u$ and parameters satisfied (\ref{6-13}). As for the first equation  of (\ref{5-5}) it gives for function $v$
\begin{equation}\label{6-15}
s^{\frac 32}v'+v^2=0.
\end{equation}
The solution of this equation is 
$$v=\frac{\sqrt s}{c\sqrt s-2} \qquad ==> \qquad g'=\frac 1{s(c\sqrt s-2)}.$$
Omitting the constant of integration we obtain the solution
$$f=0, \qquad g=\ln\Big(1-\frac 2{cr}\Big)=\ln\Big(1+\frac {r_0}{r}\Big),$$
where we introduce a new length dimensional parameter $r_0$.\\
Thus the correspondent coframe is
\begin{equation}\label{6-16}
\vt^0=dt,\qquad \vt^i=\Big(1+\frac {r_0}{r}\Big)dx^i, \qquad i=1,2,3.
\end{equation}
This coframe generates an asymptotically-flat metric
\begin{equation}\label{6-17}
ds^2=dt^2-\Big(1+\frac {r_0}{r}\Big)^2(dx^2+dy^2+dz^2).
\end{equation}
Let us first determine the sign of the parameter $r_0$. The metric (\ref{6-17}) represents a point-like solution if the correspondent ADM-mass is positive. 
Rewrite the metric (\ref{6-17})  in the spherical  Schwarzschild-type coordinates (where the circumference of a circle with center at the origin is equal to $2\pi r$).
In the spherical coordinates the metric is 
$$ds^2=dt^2-\Big(1+\frac {r_0}{r}\Big)^2(dr^2+r^2d\Omega^2).$$
Using now the translation
$$\tilde{r}= r+r_0$$ we obtain the asymptotic-flat metric in the Schwarzschild coordinates
\begin{equation}\label{6-18}
ds^2=dt^2-\frac {d\tilde{r}^2}{(1-\frac {r_0}{\tilde{r}})^2}-\tilde{r}^2d\Omega^2.
\end{equation}
The ADM mass for the metric (\ref{6-18}) takes the form
\begin{equation}\label{6-19}
m:=\lim_{\tilde{r}\to \infty}\frac {\tilde r}2\Big(1-(1-\frac {r_0}{\tilde{r}})^2\Big)=r_0.
\end{equation}
Thus by taking the parameter $r_0$ to be positive we obtain a particle-type solution with a finite positive ADM-mass. \\
The metric (\ref{6-17}) is singular at the origin of coordinates $r=0$ and consequently the metric (\ref{6-18}) is singular at $\tilde{r}=r_0$. In order to clarify the nature of this singularity compute the scalar curvature of the metric (\ref{6-17}) via the formula (\ref{6-1}). The result is  
\begin{equation}\label{6-20}
R=\frac {r_0^2}{(r_0+r)^4}=\frac {r_0^2}{\tilde r^4}.
\end{equation}
This function is non-zero and regular for all  values of $r$ including the origin ($r=0$). \\
The proper distance for a radial null geodesic in the metric (\ref{6-18}) is equal to the proper time and attach the infinity
$$
l=t=\int^{\tilde r_1}_{r_0} \frac {d\tilde r}{1-\frac {r_0}{\tilde r}} \to \infty
$$
Thus the point $r=0$ ($\tilde r=r_0$) does not belong to any final part of the space-time. 
\subsect{ \  $\boldsymbol{v=0\qquad ==> \qquad\r_3=0}$}
The equation (\ref{5-15}) has in this case a non-trivial solution only if $\r_3=0$. The system (\ref{5-5}) yields $u=0$ thus it does not has a nontrivial solution.
\subsect{Proportional solution. }
We see from the equation (\ref{5-15}) that generic solutions $u$ and $v$  satisfy the homogeneous algebraic equation thus they should be  proportional one to the other
$$u=\la v, \qquad \la=\la(\r_1,\r_3).$$
Rewrite the third equation of (\ref{5-5}) as 
$$\rho_1\Big(2\frac v{\sqrt s}+(\frac u{\sqrt s})^2+2(\frac v{\sqrt s})^2\Big)+
2\rho_3\Big(\frac u{\sqrt s}+\frac v{\sqrt s}+2\frac u{\sqrt s}\frac v{\sqrt s}+(\frac v{\sqrt s})^2\Big)=0.$$
Substituting $u=\la v$ we obtain an algebraic  linear equation for $\Big(\frac v {\sqrt s}\Big)$ with constant coefficients. Thus the general solution of the system should be of the form 
\begin{equation}\label{6-21}
u=a\sqrt s, \qquad v=b\sqrt s,
\end{equation}
where $a$ and $b$ are constants that depend  on the parameters $\r_1,\r_3.$ \\
Consequently the ansatz functions $f$ and $g$ satisfy
\footnote{By rescaling the coordinates in the coframe and in the line element we can use the same constant of integration in the functions $f$ and $g$.}  
\br\label{6-22}
f'&=&us^{-3/2}=\frac as \qquad ==>\qquad f=a\ln\frac s{r_0^2},\\
\label{6-23}
g'&=&vs^{-3/2}=\frac bs \qquad ==>\qquad g=b\ln\frac s{r_0^2}.
\er
The scalar curvature (\ref{6-1}) for this type of solutions takes the form
\begin{equation} \label{6-24}
R=-4\frac {\r_1}{\r_1+2\r_3}\cdot\frac {r_0^{4b}(a-b)^2}{s^{4b+2}}.
\end{equation}
Substituting the solutions (\ref{6-21}) in the system (\ref{5-5}) we obtain an algebraic system for two scalars $a$ and $b$:
\begin{equation}\label{6-25}
\left\{
\begin{array}{l}
\r_1(a+2ab-2b^2+a^2)+2\r_3(b+b^2)=0\\
\r_1(2b+4ab-a^2+2b^2)+2\r_3(a+b+2a^2+2ab+b^2)=0\\
\r_1(2b+a^2+2b^2)+2\r_3(a+b+2ab+b^2)=0.
\end{array}\right.
\end{equation}
Extracting the second equation from the first and from the third we obtain an equivalent system
\begin{equation}\label{6-26}
\left\{
\begin{array}{l}
\r_1(a+2ab-2b^2+a^2)+2\r_3(b+b^2)=0\\
(a+b)\Big(\r_1(2b-a)+2\r_3a\Big)=0\\
a\Big(\r_1(2b-a)+2\r_3a\Big)=0.\\
\end{array}\right.
\end{equation}
The case $a=0$ and consequently $u=0$ was treated above thus we obtain from the third equation of this system that the nontrivial solutions should satisfy 
$$\r_1(2b-a)+2\r_3a=0.$$
Rejecting the case  $\r_1= 0$ (also treated above) we obtain
 \begin{equation}\label{6-27}
b=a\frac{\r_1-2\r_3}{2\r_1}.
\end{equation}
Substituting this relation in the first equation of (\ref{6-26}) we obtain
 \begin{equation}\label{6-28}
(\r_1-\r_3)(3\r_1-2\r_3)(\r_1+2\r_3)a=2\r_1(\r_3-\r_1)(\r_1+2\r_3).
\end{equation}
In order to satisfy this equation we have to separate the following three cases:\\
$\bullet\qquad\boldsymbol{\r_1=\r_3}$.\\
Via (\ref{6-27}) the constants are connected as
$$a=-2b$$
and the system (\ref{6-26}) satisfied identically.\\
Hence the coframe 
\begin{equation}\label{6-29}
\vt^0=\Big(\frac{r}{r_0}\Big)^{2b}dt,\qquad \vt^i= \Big(\frac{r_0}{r}\Big)^{b}dx^i
\end{equation}
includes two free parameters $r_0$ and $b$. 
The correspondent line element is
\begin{equation}\label{6-30}
ds^2=\Big(\frac{r}{r_0}\Big)^{4b}dt^2-\Big(\frac{r_0}{r}\Big)^{2b}(dx^2+dy^2+dz^2).
\end{equation}

$\bullet\qquad\boldsymbol{\r_1=-2\r_3}$.\\
Via (\ref{6-27}) the constants in this case are equal
$$a=b$$
and the system (\ref{6-26}) satisfied identically.
Hence the coframe is
\begin{equation}\label{6-31}
\vt^0=\Big(\frac{r}{r_0}\Big)^{a}dt,\qquad \vt^i= \Big(\frac{r}{r_0}\Big)^{a}dx^i.
\end{equation}
The correspondent line element is conformal flat
\begin{equation}\label{6-32}
ds^2=\Big(\frac{r}{r_0}\Big)^{2a}(dt^2-dx^2-dy^2-dz^2).
\end{equation}
$\bullet\qquad\boldsymbol{3\r_1=2\r_3}$.\\
The equation (\ref{6-28}) does not satisfied for any finite value of $a$ thus 
the field equation does not has a non-trivial solution.\\
$\bullet\qquad\boldsymbol{Algebraic \ general \ case}$.\\
Consider the generic case 
$$
\r_1\ne\r_3, \qquad \r_1\ne -2\r_3, \qquad 3\r_1\ne 2\r_3.
$$
From the equations (\ref{6-28} - \ref{6-27}) the value of the constants are
\begin{equation}\label{6-33}
a=\frac{2\r_1}{2\r_3-3\r_1}, \qquad b=\frac{\r_1-2\r_3}{2\r_3-3\r_1}.
\end{equation}
Now we derive by integration the ansatz functions 
\begin{equation}\label{6-34}
f=a\ln{\frac s{r_0^2}}, \qquad g=b\ln{\frac s{r_0^2}}.
\end{equation}
Thus the coframe is 
\begin{equation}\label{6-35}
\vt^0=\Big(\frac r{r_0}\Big)^{2a}dt, \qquad \vt^i=\Big(\frac r{r_0}\Big)^{2b}dx^i,\qquad i=1,2,3.
\end{equation}
The correspondent metric element is
\begin{equation}\label{6-36}
ds^2=\Big(\frac r{r_0}\Big)^{4a}dt^2-\Big(\frac r{r_0}\Big)^{4b}(dx^2+dy^2+dz^2).
\end{equation}
Note that the parameters $a$ and $b$ are completely defined by Eq. (\ref{6-33}).\\
Consider now how these proportional solutions related to the special solutions considered above.
\\$\boldsymbol{1) \qquad \qquad \r_1=0}$\\
We obtain from the equations (\ref{6-33})
$$a=0,\qquad b=-1$$
thus the solution (\ref{6-35}) takes the form
$$\vt^0=dt, \qquad \vt^i=\Big(\frac {r_0}{r}\Big)^{2}dx^i,\qquad i=1,2,3.$$
The correspondent metric is 
$$
ds^2=dt^2-=\Big(\frac {r_0}{r}\Big)^{2}(dr^2+r^2d\Omega^2)
$$
Using the  following change  of the radial coordinate 
$$
r=\frac{r_0^2}{\tilde r}
$$
we obtain the Minkowskian metric.
Thus the  Schwarzschild coframe  yields a unique non-flat metric. 
\\ $\boldsymbol{2) \qquad \qquad \r_1=\r_3}$\\
Substating this relain in (\ref{6-33})
we obtain
$$a=-2,\qquad b=1$$
and the solution is 
$$\vt^0=\Big(\frac r{r_0}\Big)^{-4}dt, \qquad \vt^i=\Big(\frac r{r_0}\Big)^{2}dx^i,\qquad i=1,2,3.$$
This is a special case of (\ref{6-10}).
\\ $\boldsymbol{3) \qquad \qquad \r_1+\r_3=0}$\\
The relation (\ref{6-33}) gives
$$a=-\frac 25,\qquad b=-\frac 35$$
and the solution (\ref{6-35}) is 
$$\vt^0=\Big(\frac{r_0} r\Big)^{\frac 45}dt, \qquad \vt^i=\Big(\frac {r_0}r\Big)^{\frac 65}dx^i,\qquad i=1,2,3.$$
Thus the field equation has two non-trivial solutions. 
\\ $\boldsymbol{3) \qquad \qquad \r_3=0}$\\
The parameters of the cofrane are
$$a=-\frac 23, \qquad b=-\frac 13$$
and the solution is 
$$\vt^0=\Big(\frac {r_0}r\Big)^{\frac 43}dt, \qquad \vt^i=\Big(\frac {r_0}r\Big)^{\frac 23}dx^i,\qquad i=1,2,3.$$
Thus for $\r_3=0$ the field equation has an unique solution.
\\ $\boldsymbol{4) \qquad \qquad \r_1=-2\r_3}$\\
The parameters take the values 
$$a=-\frac 12,\qquad b=-\frac 12$$ 
and the solution is 
$$\vt^0=\frac {r_0}{r}dt, \qquad \vt^i=\frac {r_0}{r}dx^i,\qquad i=1,2,3.$$
This is a special case of the conformal flat solution (\ref{6-31}).
\\ $\boldsymbol{5)\qquad \qquad \r_1=2\r_3}$\\
The parameters $a,b$ are non-defined and the proportional solution does not exist.\\
The results of our analyse can be stated in the following theorem:
\thm 
The field equation (\ref{3-3}) has a non-trivial sphrically symmetric static solytion of the ``diogonal'' type for all values of the free parameters $\r_1,\r_3$ exept of the case $$\r_1=2\r_3.$$
For 
$$\r_1=\r_3$$
exists a family of solutions, which depends on an arbitrary function of radial distance.\\
For 
$$\r_1=-2\r_3$$
exists a family af conformal flat solutions, which parameterized by an arbitrary constant.\\
For $$\r_1+\r_3=0$$ 
exist two non-trivial solutions.
For all remain values of the parameters $\r_1,\r_3$ the solution is unique.\\
The assymptotyc flat solution exist only for
$$\r_1=0 \qquad \text{or}\qquad \r_1+\r_3=0.$$
A unique solution wich gives the Newtonian limit exists for $$ \r_1=0$$
and leads to the Schwarzschild metric.
\ethm 
\sect{Metrics generated by proportional solutions. }
In the previous section it was shown that for generic values of free parameters of the theory the field equation (\ref{3-3}) has a unique non-trivial static spherical-symmetric coframe solution of the form (\ref{6-35}). This solution generates the metric (\ref{6-36}). The physical meaning of these metrics require consideration of the topological properties of the manifold where the metrics can be realized. \\
The camponents of the metric tensor (\ref{6-36}) have two  coordinate singularities: one  at the origin of coordinates $r=0$ and the second  at the coordinate infinity $r=\infty$. 
In order to clearify the nature of these singularities calculate the scalar curvature of the metric  (\ref{6-36}) Using the formula (\ref{6-1}) we obtain
\begin{equation}\label{7-1}
R=-\frac{4\r_1}{\r_1+2\r_3}\frac{(a-b)^2}{r_0^2}\Big(\frac{r_0}{r}\Big)^{4b+2}
\end{equation}
Thus in the case 
\begin{equation}\label{7-2}
2b+1>0 \qquad<==> \qquad \frac {\r_1}{\r_3}<-2 \quad \text{or} \quad \frac {\r_1}{\r_3}>\frac 23
\end{equation}
the scalar curvature in the origin $r=0$ is infinity.\\
Accordingly in the case 
\begin{equation}\label{7-3}
2b+1>0 \qquad<==> \qquad -2<\frac {\r_1}{\r_3}<\frac 23
\end{equation}
the scalar curvature at the second coordinate singularity  $r=\infty$ is infinity.\\
Note also the sign of the scalar curvature:
\begin{equation}\label{7-4}
R<0 \qquad<==> \qquad -2<\frac {\r_1}{\r_3}<0
\end{equation}
and
\begin{equation}\label{7-5}
R>0 \qquad<==> \qquad \frac {\r_1}{\r_3}>0 \quad \text{or} \quad \frac{\r_1}{\r_3}<-2.
\end{equation}
Let us transform now the metric (\ref{6-36}) to the Schwarzschild coordinates. In the spherical coordinates this metric takes the form
\begin{equation}\label{7-6}
ds^2=\Big(\frac r{r_0}\Big)^{4a}dt^2-\Big(\frac r{r_0}\Big)^{4b}(dr^2+r^2d\Omega^2).
\end{equation}
In the case 
\begin{equation}\label{7-7}
2b+1\ne0 \qquad<==> \qquad \frac {\r_1}{\r_3}\ne-2 
\end{equation}
we can use the transformation
\begin{equation}\label{7-8}
\tilde r=r\Big(\frac r{r_0}\Big)^{2b} \qquad<==> \qquad r=r_0\Big(\frac {\tilde r}{r_0}\Big)^{\frac 1{2b+1}}
\end{equation}
The metric (\ref{6-36}) in the new (Schwarzschild) coordinates takes the form
\begin{equation}\label{7-9}
ds^2=\Big(\frac {\tilde r}{r_0}\Big)^{\frac {4a}{2b+1}}dt^2-\frac {dr^2}{(2b+1)^2}-r^2d\Omega^2.
\end{equation}
Note that that the hyper-spaces $t=const$ is not flat, although $b$ is constant. The four dimensional metric (\ref{7-9}) is spherically symmetric thus for analyzing the nature of the singularity it is enough to consider it's two-dimensional part ``$\tilde r - t$''. The proper radial distance from an arbitrary point $r_1$ to a singular point $r_*$ is 
\begin{equation}\label{7-10} 
l=\Big|\int_{\tilde{r}_1}^{\tilde{r}_*}\frac {d\tilde{r}}{2b+1}\Big|=\Big|\frac{\tilde{r}_*-\tilde{r}_1}{2b+1}\Big|.
\end{equation}
As for a proper time to attach the singular point $r_*$ by a radial null geodesic is
\begin{equation}\label{7-11}  
t=\Big|\frac 1{2b+1}\int_{\tilde{r}_1}^{\tilde{r}_*}\Big(\frac {r_0}{\tilde r}\Big)^{\frac {2a}{2b+1}}d\tilde {r}\Big|=\frac {|r_*^m-r_1^m|} {|2b-2a+1|},
\end{equation}
where 
$$m=\frac{2b-2a+1}{2b+1}=\frac{2\r_3-3\r_1}{\r_1+2\r_3}.$$
Thus we obtain the following regimes for the singularity in the origin of the coordinates:
\begin{center}
%
\begin{tabular}{|c|c|c|c|c|}
\hline
 & &  &  &\\ 
$\boldsymbol{r=0}$&$\frac {\r_1}{\r_3}<-2$&$-2<\frac {\r_1}{\r_3}<-\frac 12$&$-\frac 12<\frac {\r_1}{\r_3}<\frac 23$&$\frac {\r_1}{\r_3}>\frac 23$\\ 
  & &  &  &\\ 
\hline
New coordinates&$\tilde r=0$&$\tilde r=\infty$&$\tilde r=\infty$&$\tilde r=0$\\
\hline
Scalar curvature&$R=\infty$&$R=0$&$R=0$&$R=\infty$\\ 
\hline
Proper distance&finite&$\infty$&$\infty$&finite\\ 
\hline
Parameter $m$  & $m<0$ & $m>0$ & $m<0$ &$m>0$\\ 
\hline
Proper time    & $\infty$ & $\infty$ & finite & $\infty$\\ 
\hline
\end{tabular}
\end{center}
Analyses of this values gives the following regimes:
\begin{itemize}
\item \qquad \qquad\qquad  $\boldsymbol{\frac {\r_1}{\r_3} < -2}$ and $\boldsymbol{\frac {\r_1}{\r_3}>\frac 23}$\\
For these ranges of the parameters  we obtain that in the point $r=0$ the scalar curvature is singular. This point is located at a finite proper distance from an arbitrary point on the manifold but it takes an infinite time to reach this point by a radial null geodesic. Therefore for these values of the free parameters the origin of the coordinates is a physical singular point of the same type as  the central point of the Schwarzschild metric.  
\item \qquad \qquad \qquad $\boldsymbol{-2<\frac {\r_1}{\r_3}<-\frac 12}$\\
The scalar curvature in the origin of coordinates is zero. This point is located on an infinite proper distance from an arbitrary point on the manifold and it takes an infinity time to reach this point by a null geodesic. Therefore in this case the point $r=0$ does not belong to any final portion of the space time.
\item\qquad \qquad \qquad $\boldsymbol{-\frac 12<\frac {\r_1}{\r_3}<\frac 23}$\\
The origin $r=0$ is a point with zero scalar curvature. It is located in an infinity distance from every point but it can be reached by a final time. Thus the singular point  is located outside of the null cone.  
\end{itemize}
As for the second coordinate singularity $r=\infty$ we have a following table of regimes:

\begin{center}
%
%
\begin{tabular}{|c|c|c|c|c|}
\hline
 & &  &  &\\ 
$\boldsymbol{r=\infty}$&$\frac {\r_1}{\r_3}<-2$&$-2<\frac {\r_1}{\r_3}<-\frac 12$&$-\frac 12<\frac {\r_1}{\r_3}<\frac 23$&$\frac {\r_1}{\r_3}>\frac 23$\\ 
  & &  &  &\\ 
\hline
New coordinates&$\tilde r=\infty$&$\tilde r=0$&$\tilde r=0$&$\tilde r=\infty$\\
\hline
Scalar curvature&$R=0$&$R=\infty$&$R=\infty$&$R=0$\\ 
\hline
Proper distance&$\infty$&finite&finite&$\infty$\\ 
\hline
Parameter $m$  & $m<0$ & $m>0$ & $m<0$ &$m>0$\\ 
\hline
Proper time    & finite & finite &$\infty $ & $\infty$\\ 
\hline
\end{tabular}
\end{center}
Thus we have the following regimes:
\begin{itemize}
\item \qquad \qquad\qquad  $\boldsymbol{\frac {\r_1}{\r_3} < -2}$\\
The scalar singularity in the point $r=\infty$ is zero. This point is located on an infinity distance from every finite point on the manifold. Although it takes a finite time to reach this point from an arbitrary point on the manifold by a radial null geodesic. Thus the point $r=\infty$ is located outside of the null cone.
\item \qquad \qquad\qquad  $\boldsymbol{-2<\frac {\r_1}{\r_3} < -\frac 12}$\\
The scalar curvature is infinite thus the point $r=\infty$ is a physical singularity. This point is located on a finite distance from an arbitrary point and can be reached by a finite time. The singularity is similar to a vertex of the cone - conical singularity.
\item \qquad \qquad\qquad  $\boldsymbol{-\frac 12<\frac {\r_1}{\r_3} < \frac 23}$\\ 
For these parameters the point $r=\infty$ is, also, a physical singularity. It can be however be reached by a infinite time nevertheless it is located on a  finite proper distance. Thus this singularity is similar to the central point  of the Schwarzschild metric.
\item \qquad \qquad\qquad\qquad\quad  $\boldsymbol{\frac {\r_1}{\r_3} > \frac 23}$\\ 
The proper distance and the proper time is infinite. Therefore the point is located outside of any finite area of space-time.
\end{itemize}
\sect{Conclusions and discussion. }
We study the general quadratic Lagrangian of pure teleparallel theory. This Lagrangian is a linear combination of three covariant and global $SO(1,3)$ invariant terms with free dimensionless coefficients $\r_1,\r_2,\r_3$. The field equations of the theory is studied by a ``diagonal'' ansatz which is a subclass of a general spherical-symmetric Einstein-Mayer ansatz. The restriction is taken for simplification the calculation. Thus we study a Lagrangian that depends only on two free parameters $\r_1,\r_3$.\\
 We obtain a formula for scalar curvature of a pseudo-Riemannian manifold with a metric constructed from the solution of the field equation. This formula shows that the sign of the scalar curvature depends only on  a relation between the parameters $\r_1$ and $\r_3$. The scalar curvature vanishes only in a case $r_1=0$ which corresponds to the teleparallel equivalent of GR.\\
We obtain exact solutions for all possible values of free parameters $\r_1,\r_3$. It is shown that the unique solution with a Newtonian limit is the Schwarzschild solution. Thus the Yang-Mills-type term of the general quadratic Lagrangian should be rejected.\\
Note that all the results valid only in the case of a restricted ``diagonal'' ansatz. Thus it is important to obtain solutions for following questions:
\begin{itemize}
\item  Which physical sense can be given to the various exact solutions?
\item How the results depends on the non-diagonal terms?
\item How the scalar curvature inequalities can be generalized to the case of three free parameters   $\r_1,\r_2,\r_3$?
\end{itemize}

\section*{Acknowledgments}
I am deeply grateful to F.W. Hehl and to S. Kaniel for constant support, interesting discussions and valuable comments.
\newpage
\appendix
\footnotesize{
\sect{Calculations with the diagonal ansatz}
We study the field equation (\ref{3-3}) with the diagonal coframe ansatz 
$$
\vt^0=e^fdt, \qquad \vt^i=e^gdx^i, \qquad i=1,2,3.
$$ 
We will use the following notations
\br
&&(\nabla f\cdot\nabla g)=f_1g_1+f_2g_2+f_3g_3\\
&&(\nabla f)^2=f_1^2+f_2^2+f_3^2\\
&&\triangle f=f_{11}+f_{22}+f_{33}
\er
Calculating the first derivatives of the diagonal coframe  we obtain \\[0.25in]
\begin{tabular}{|l|l|}
\hline
&\\
$d\vt^0=-e^{-g}(f_1\vt^{01}+f_2\vt^{02}+f_3\vt^{03})$&
$*d\vt^0=-e^{-g}(f_1\vt^{23}-f_2\vt^{13}+f_3\vt^{12})$\\  & \\
\hline & \\
$d\vt^1=-e^{-g}(g_2\vt^{12}+g_3\vt^{13})$&
$*d\vt^1=e^{-g}(g_2\vt^{03}-g_3\vt^{02})$\\ &  \\
\hline & \\
$d\vt^2=e^{-g}(g_1\vt^{12}-g_3\vt^{23})$&
$*d\vt^2=-e^{-g}(g_1\vt^{03}-g_3\vt^{01})$\\  & \\
\hline & \\
$d\vt^3=e^{-g}(g_1\vt^{13}+g_2\vt^{23})$&
$*d\vt^1=e^{-g}(g_1\vt^{02}-g_2\vt^{01})$\\ & \\
\hline
\end{tabular}
\subsect{Temporal components}
The first leading term of (\ref{3-3}) takes the form
\br
{}^{(1)}L_0&=&d*d\vt_0
=-e^{-2g}\Big(\triangle f+(\nabla f\cdot\nabla g)\Big)\vt^{123},
\er
Thus
\begin{equation}
\boxed{{}^{(1)}L_0=-e^{-2g}\Big(\triangle f+(\nabla f\cdot\nabla g)\Big)*\vt^0}
\end{equation}
As for the second leading term of (\ref{3-3})
\br
{}^{(3)}L_0 \eqq \vt_b \wedge d*( \vt_0 \wedge d \vt^b )=
-\vt^1\wedge d*(\vt^0\wedge d\vt^1)-\vt^2\wedge d*(\vt^0\wedge d\vt^2)-
\vt^3\wedge d*(\vt^0\wedge d\vt^3)\nonumber\\
\eqq -\vt^1\wedge d*(-e^{-g})(g_2\vt^{012}+g_3\vt^{013})-\vt^2\wedge d*(e^{-g})(g_1\vt^{012}-g_3\vt^{023})\nonumber\\
\eee -\vt^3\wedge d*(e^{-g})(g_1\vt^{013}+g_2\vt^{023})\nonumber\\
\eqq \vt^1\wedge d(e^{-g})(g_2\vt^{3}-g_3\vt^{2})-\vt^2\wedge d(e^{-g})(g_1\vt^{3}-g_3\vt^{1})+\vt^3\wedge d(e^{-g})(g_1\vt^{2}-g_2\vt^{1})\nonumber\\
\eqq \vt^1\wedge d(g_2dz-g_3dy)-\vt^2\wedge d(g_1dz-g_3dx)+\vt^3\wedge d(g_1dy-g_2dx)
=2e^{-2g}\triangle g \vt^{123}
\er
Thus
\begin{equation}
\boxed{{}^{(3)}L_0=2e^{-2g}\triangle g *\vt^0}
\end{equation}
The first quadratic term is
\br
{}^{(1)}Q_0 \eqq e_0 \hook( d\vt^b \wedge * d\vt_b)= e_0 \hook\Big( d\vt^0 \wedge * d\vt^0-d\vt^1 \wedge * d\vt^1-d\vt^2 \wedge * d\vt^2-d\vt^3 \wedge * d\vt^3\Big)\nonumber\\
\eqq e^{-2g}e_0 \hook\Big((f_1\vt^{01}+f_2\vt^{02}+f_3\vt^{03})\wedge 
(f_1\vt^{23}-f_2\vt^{13}+f_3\vt^{12})-(g_2\vt^{12}+g_3\vt^{13})(-g_2\vt^{03}+g_3\vt^{02})+\nonumber\\ \eee
(g_1\vt^{12}+g_3\vt^{23})(-g_1\vt^{03}-g_3\vt^{01})
-(g_1\vt^{13}+g_2\vt^{23})(g_1\vt^{02}-g_2\vt^{01})
\nonumber\\
\eqq e^{-2g}\Big(\nabla^2f+2\nabla^2 g\Big)\vt^{123}
\er
Thus
\begin{equation}
\boxed{{}^{(1)}Q_0= e^{-2g}\Big((\nabla f)^2+2(\nabla g)^2\Big)*\vt^0}
\end{equation}
The second quadratic term is 
\br
{}^{(2)}Q_0 \eqq(e_0\hook d\vt^b)\wedge*d\vt_b=(e_0\hook d\vt^0)\wedge*d\vt_0\nonumber\\
\eqq e^{-2g}(f_1\vt^{1}+f_2\vt^{2}+f_3\vt^{3})\wedge 
(f_1\vt^{23}-f_2\vt^{13}+f_3\vt^{12})= e^{-2g}\nabla^2f\vt^{123}
\er 
Thus
\begin{equation}
\boxed{{}^{(2)}Q_0= e^{-2g}(\nabla f)^2*\vt^0}
\end{equation}
The next non-zero quadratic term is
\br
{}^{(6)}Q_0 \eqq d\vt_b\wedge*(\vt_0\wedge d\vt^b)=-d\vt^1\wedge*(\vt^0\wedge d\vt^1)-d\vt^2\wedge*(\vt^0\wedge d\vt^2)-d\vt^3\wedge*(\vt^0\wedge d\vt^3)\nonumber\\
\eqq -e^{-2g}\Big((g_2\vt^{12}+g_3\vt^{13})\wedge*(g_2\vt^{012}+g_3\vt^{013})+(g_1\vt^{12}-g_3\vt^{23})\wedge*(g_1\vt^{012}-g_3\vt^{023})\nonumber\\
\eee
\qquad\qquad +(g_1\vt^{13}+g_2\vt^{23})\wedge*(g_1\vt^{013}+g_2\vt^{023})\Big)\nonumber\\
\eqq -e^{-2g}\Big((g_2\vt^{12}+g_3\vt^{13})\wedge(g_2\vt^{3}-g_3\vt^{2})+(g_1\vt^{12}-g_3\vt^{23})\wedge(g_1\vt^{3}-g_3\vt^{1})\nonumber\\
\eee
\qquad\qquad +(g_1\vt^{13}+g_2\vt^{23})\wedge(-g_1\vt^{2}+g_2\vt^{1})\Big)= -2e^{-2g}\nabla^2g\vt^{123}
\er 
Thus
\begin{equation}
\boxed{{}^{(6)}Q_0= -2e^{-2g}(\nabla g)^2*\vt^0}
\end{equation}
As for the next non-zero quadratic term ${}^{(7)}Q_0$ we have 
\br
{}^{(7)}Q_0 \eqq e_0\hook\Big(\vt_c\wedge d\vt^b\wedge*(d\vt^c\wedge\vt_b)\Big)\nonumber\\
\eqq 2e_0\hook\Big(-\vt^0\wedge d\vt^1\wedge *(d\vt^0\wedge\vt^1)-\vt^0\wedge d\vt^2\wedge *(d\vt^0\wedge\vt^2)-\vt^0\wedge d\vt^3\wedge *(d\vt^0\wedge\vt^3)\nonumber\\
\eee \qquad +\vt^1\wedge d\vt^2\wedge *(d\vt^1\wedge\vt^2)+\vt^1\wedge d\vt^3\wedge *(d\vt^1\wedge\vt^3)+\vt^2\wedge d\vt^3\wedge *(d\vt^2\wedge\vt^3)\Big)
\nonumber\\
\eqq 2e^{-2g}e_0\hook\Big((g_2\vt^{012}+g_3\vt^{013})\wedge(f_2\vt^3-f_3\vt^2)+(g_1\vt^{012}-g_3\vt^{023})\wedge(f_1\vt^3-f_3\vt^1)\nonumber\\
\eee  
+(g_1\vt^{013}+g_2\vt^{023})\wedge(-f_1\vt^2+f_2\vt^1) -g_3\vt^{123}\wedge g_3\vt^0-g_2\vt^{123}\wedge g_2\vt^0-g_1\vt^{123}\wedge g_1\vt^0\Big)\nonumber\\
\eqq 2e^{-2g}(2\nabla f\cdot\nabla g+\nabla^2 g)\vt^{123}
\er
Thus
\begin{equation}
\boxed{{}^{(7)}Q_0= 2e^{-2g}\Big(2(\nabla f\cdot\nabla g)+(\nabla g)^2 \Big)*\vt^0}
\end{equation}
The last non-zero quadratic term ${}^{(8)}Q_0$ is
\br
{}^{(8)}Q_0\eqq(e_0\hook d\vt^b)\wedge\vt_c\wedge*(d\vt^c\wedge\vt_b )\nonumber\\\eqq
-(e_0\hook d\vt^0)\wedge\Big(\vt^1\wedge*(d\vt^1\wedge\vt^0)+\vt^2\wedge*(d\vt^2\wedge\vt^0)+\vt^3\wedge*(d\vt^3\wedge\vt^0)\Big)\nonumber\\
\eqq -e^{-2g}(f_1\vt^1+f_2\vt^2+f_3\vt^3)\wedge\Big(\vt^1\wedge*(g_2\vt^{012}+g_3\vt^{013})-\vt^2\wedge*(g_1\vt^{012}-g_3\vt^{023})-\nonumber\\
\eee \qquad \vt^3\wedge*(g_1\vt^{013}+g_2\vt^{023})\Big)\nonumber\\
\eqq -e^{-2g}(f_1\vt^1+f_2\vt^2+f_3\vt^3)\wedge\Big(g_2\vt^{13}-g_3\vt^{12}-g_1\vt^{23}+g_3\vt^{21}+g_1\vt^{32}-g_2\vt^{31}\Big)\nonumber\\
\eqq 2e^{-2g}(\nabla f\cdot\nabla g)\vt^{123}
\er
Thus
\begin{equation}
\boxed{{}^{(8)}Q_0= 2e^{-2g}(\nabla f\cdot\nabla g)*\vt^0}
\end{equation}
\subsect{Spatial components}
The first leading term takes the form
\br
{}^{(1)}L_1\eqq d*d\vt_1=d\Big(e^f(-g_2dt\wedge dz+g_3dt\wedge dy)\Big)\nonumber\\\eqq -e^{-2g}\Big((g_{12}+f_1g_2)\vt^{013}+(g_{22}+g_{33}+f_2g_2+f_3g_3)\vt^{023}-(g_{13}+f_1g_3)\vt^{012}\Big)\nonumber\\\eqq
e^{-2g}\Big((\triangle g+\nabla f\cdot\nabla g-g_{11}-f_1g_1)*\vt^{1}-(g_{12}+f_1g_2)*\vt^{2}-(g_{13}+f_1g_3)*\vt^{3}\Big)
\er
Thus
\begin{equation}
\boxed{{}^{(1)}L_i=-e^{-2g}\bigg(\Big(\triangle g+(\nabla f\cdot\nabla g)\Big)\eta_{ik}+g_{ik}+f_ig_k\bigg)*\vt^k}
\end{equation}
The second leading term is
\br
{}^{(3)}L_1\eqq \vt_b\wedge d*(\vt_1\wedge d\vt^b)\nonumber\\\eqq
-\vt^0\wedge d*(\vt^1\wedge d\vt^0)+\vt^2\wedge d*(\vt^1\wedge d\vt^2)+
\vt^3\wedge d*(\vt^1\wedge d\vt^3)\nonumber\\ 
\eqq -\vt^0\wedge d*(e^{-g})(f_2\vt^{012}+f_3\vt^{013})
-\vt^2\wedge d*(e^{-g})g_3\vt^{123})+\vt^3\wedge d*(e^{-g})g_2\vt^{123})\nonumber\\ 
\eqq -\vt^0\wedge d(e^{-g})(f_2\vt^{3}-f_3\vt^{2})
-\vt^2\wedge d(e^{-g}g_3\vt^{0})+\vt^3\wedge d(e^{-g}g_2\vt^{0})\nonumber\\ 
\eqq -\vt^0\wedge d(f_2dz-f_3dy)
-\vt^2\wedge d(e^{f-g}g_3dt)
+\vt^3\wedge d(e^{f-g}g_2dt)\nonumber\\ 
\eqq -\vt^0\wedge \Big((f_{22}+f_{33})dy\wedge dz+f_{12}dx\wedge dz -f_{13}dx\wedge dy\Big)\nonumber\\ 
\eee
-\vt^2\wedge e^{f-g}\Big([g_{13}+(f_1-g_1)g_3]dx\wedge dt+[g_{33}+(f_3-g_3)g_3]dz\wedge dt\Big)\nonumber\\ 
\eee
+\vt^3\wedge e^{f-g}\Big([g_{12}+(f_1-g_1)g_2]dx\wedge dt+[g_{22}+(f_2-g_2)g_2]dy\wedge dt\Big)
\nonumber\\ 
\eqq
e^{-2g}\Big((-\triangle f-\triangle g+(\nabla g)^2-\nabla f\cdot\nabla g+f_{11}+g_{11}-g_1^2+f_1g_1)*\vt^{1}\nonumber\\
\eee +(f_{12}+g_{12}+f_1g_2-g_1g_2)*\vt^{2}+(f_{13}+g_{13}+f_1g_3-g_1g_3)*\vt^{3}\Big)
\er
Thus 
\begin{equation}
\boxed{{}^{(3)}L_i= 
e^{-2g}\bigg(\Big(\triangle f+\triangle g-(\nabla g)^2+
(\nabla f\cdot\nabla g)\Big)\eta_{ik}+f_{ik}+g_{ik}-g_ig_k+f_ig_k\bigg)*\vt^k}
\end{equation}
Using the calculations of ${}^{(7)}Q_0$ we obtain the first quadratic term as  
\begin{equation}
\boxed{{}^{(1)}Q_i = e_1 \hook( d\vt^b \wedge * d\vt_b)=e^{-2g}\Big((\nabla f)^2+2(\nabla g)^2 \Big)\eta_{ik}*\vt^k}
\end{equation}
As for the second quadratic term 
\br
{}^{(2)}Q_1 \eqq(e_1\hook d\vt^b)\wedge*d\vt_b=(e_1\hook d\vt^0)\wedge*d\vt^0-
(e_1\hook d\vt^1)\wedge*d\vt^1-(e_1\hook d\vt^2)\wedge*d\vt^2-(e_1\hook d\vt^3)\wedge*d\vt^3\nonumber\\ 
\eqq e^{-2g}\Big(-\vt^0\wedge(f_1\vt^{23}-f_2\vt^{13}+f_3\vt^{12})-(g_2\vt^2+g_3\vt^3)\wedge (-g_2\vt^{03}+g_3\vt^{02})\nonumber\\ 
\eee-g_1\vt^2\wedge (g_1\vt^{03}-g_3\vt^{01})-g_1\vt^3\wedge(g_1\vt{02}-g_2\vt^{01})\Big)\nonumber\\ 
\eqq e^{-2g}\Big(-(f_1^2+g_2^2+g_3^2+2g_1^2)\vt^{023}+(f_1f_2+g_1g_2)\vt^{013}-
(f_1f_3+g_1g_3)\vt^{012}\Big)
\er
Thus
\begin{equation}
\boxed{{}^{(2)}Q_i =e^{-2g}\Big((\nabla g)^2\eta_{ik}-(f_if_k+g_ig_k)\Big)*\vt^k}
\end{equation}
The next non-zero quadratic term is
\br
{}^{(6)}Q_1 \eqq d\vt_b\wedge*(\vt_1\wedge d\vt^b)=
-d\vt^0\wedge*(\vt^1\wedge d\vt^0)+d\vt^2\wedge*(\vt^1\wedge d\vt^2)+d\vt^3\wedge*(\vt^1\wedge d\vt^3)\nonumber\\ 
\eqq e^{-2g}\Big((f_1\vt^{01}+f_2\vt^{02}+f_3\vt^{03})\wedge *(f_2\vt^{012}+f_3\vt^{013})-\nonumber\\
\eee \qquad\qquad
(g_1\vt^{12}-g_3\vt^{23})\wedge *g_3\vt^{123}+(g_1\vt^{13}+g_2\vt^{23})\wedge *g_2\vt^{123}\Big)\nonumber\\ 
\eqq e^{-2g}\Big((f_1\vt^{01}+f_2\vt^{02}+f_3\vt^{03})\wedge (f_2\vt^{3}-f_3\vt^{2})-(g_1\vt^{12}-g_3\vt^{23})\wedge g_3\vt^{0}+(g_1\vt^{13}+g_2\vt^{23})\wedge *g_2\vt^{0}\Big)\nonumber\\ 
\eqq e^{-2g}\Big((f_2^2+f_3^2+g_2^2+g_3^2)\vt^{023}+(f_1f_2+g_1g_2)\vt^{013}-(f_1f_3+g_1g_3)\vt^{012}\Big)
\er
Thus
\begin{equation}
\boxed{{}^{(6)}Q_i =-e^{-2g}\bigg(\Big((\nabla f)^2+(\nabla g)^2\Big)\eta_{ik}+f_if_k+g_ig_k\bigg)*\vt^k}
\end{equation}
Using the calculations above for ${}^{(7)}Q_0$ we obtain for the next non-zero quadratic term
\brn
{}^{(7)}Q_1 \eqq e_1\hook\Big(\vt_c\wedge d\vt^b\wedge*(d\vt^c\wedge\vt_b)\Big)=-2e^{-2g}(2\nabla f\cdot\nabla g+\nabla^2 g)*\vt^1
\ern
Thus
 \begin{equation}
\boxed{{}^{(7)}Q_i =2e^{-2g}\Big(2(\nabla f\cdot\nabla g)+(\nabla g)^2 \Big)\eta_{ik}*\vt^k}
\end{equation}
As for the last non-zero quadratic term 
\brn
{}^{(8)}Q_0\eqq(e_1\hook d\vt^b)\wedge\vt_c\wedge*(d\vt^c\wedge\vt_b )\\ 
\eqq -(e_1\hook d\vt^0)\wedge\Big(\vt^1\wedge *(d\vt^1\wedge\vt^0)+\vt^2\wedge *(d\vt^2\wedge\vt^0)+\vt^3\wedge *(d\vt^3\wedge\vt^0)\Big)\\ 
\eee +(e_1\hook d\vt^1)\wedge\Big(-\vt^0\wedge *(d\vt^0\wedge\vt^1)+\vt^2\wedge *(d\vt^2\wedge\vt^1)+\vt^3\wedge *(d\vt^3\wedge\vt^1)\Big)\\ 
\eee +(e_1\hook d\vt^2)\wedge\Big(-\vt^0\wedge *(d\vt^0\wedge\vt^2)+\vt^1\wedge *(d\vt^1\wedge\vt^2)+\vt^3\wedge *(d\vt^3\wedge\vt^2)\Big)\\ 
\eee +(e_1\hook d\vt^3)\wedge\Big(-\vt^0\wedge *(d\vt^0\wedge\vt^3)+\vt^1\wedge *(d\vt^1\wedge\vt^3)+\vt^2\wedge *(d\vt^2\wedge\vt^3)\Big)\\
\eqq e^{-2g}\bigg(
f_1\vt^0\wedge 
\Big(\vt^1\wedge *(g_2\vt^{012}+g_3\vt^{013})-\vt^2\wedge *(g_1\vt^{012}-g_3\vt^{023})-\vt^3\wedge *(g_1\vt^{013}+g_2\vt^{023})\Big)\\ 
\eee \qquad\quad +(g_2\vt^2+g_3\vt^3)\wedge\Big(\vt^0\wedge*(f_2\vt^{012}+f_3\vt^{013})+\vt^2\wedge*g_3\vt^{123}+\vt^3\wedge*g_2\vt^{123}\Big)\\
 \eee \qquad \quad +g_1\vt^2\wedge \Big(\vt^0\wedge *(f_1\vt^{012}-f_3\vt^{023})+\vt^1\wedge *g_3\vt^{123}-\vt^3\wedge *g_3\vt^{123}\Big)\\
 \eee \qquad\quad +g_1\vt^3\wedge \Big(\vt^0\wedge *(f_1\vt^{013}+f_2\vt^{023})-\vt^1\wedge *g_2\vt^{123}+\vt^2\wedge *g_1\vt^{123}\Big)\bigg)\\
\eqq e^{-2g}\bigg(
f_1\vt^0\wedge \Big(\vt^1\wedge (g_2\vt^{3}-g_3\vt^{2})-\vt^2\wedge(g_1\vt^{3}-g_3\vt^{1})+\vt^3\wedge (g_1\vt^{2}-g_2\vt^{1})\Big)\\ 
\eee \qquad\quad +(g_2\vt^2+g_3\vt^3)\wedge\Big(\vt^0\wedge(f_2\vt^{3}-f_3\vt^{2})+\vt^2\wedge g_3\vt^{0}+\vt^3\wedge g_2\vt^{0}\Big)\\
 \eee \qquad \quad +g_1\vt^2\wedge \Big(\vt^0\wedge (f_1\vt^{3}-f_3\vt^{1})+\vt^1\wedge g_3\vt^{0}-\vt^3\wedge g_3\vt^{0}\Big)\\
 \eee \qquad\quad +g_1\vt^3\wedge \Big(\vt^0\wedge (-f_1\vt^{2}+f_2\vt^{1})-\vt^1\wedge g_2\vt^{0}+\vt^2\wedge g_1\vt^{0}\Big)\bigg)\\
\eqq e^{-2g}\Big(-(4f_1g_1+f_2g_2+f_3g_3+2g_1^2+g_2^2+g_3^2)\vt^{023}+
(2f_1g_2+g_1f_2+g_1g_2)\vt^{013}\\
\eee \qquad\quad -(2f_1g_3+g_1f_3+g_1g_3)\vt^{012}\Big)
\ern
Thus
\begin{equation}
\boxed{{}^{(8)}Q_i =e^{-2g}\bigg(\Big((\nabla f\cdot\nabla g)+(\nabla g)^2\Big)\eta_{ik}-2f_ig_k-g_if_k-g_ig_k\bigg)*\vt^k}
\end{equation}

\end{document}